\documentclass[ %
 reprint,
 amsmath,amssymb,
 aps,
aps]{revtex4-2}

\usepackage{natbib}
\usepackage{array}
\newcolumntype{P}[1]{>{\centering\arraybackslash}p{#1}}
\usepackage{graphicx}
\usepackage{dcolumn}
\usepackage{bm}

\begin{document}

\preprint{APS/123-QED}

\title{Testing Multipole Moments of compact objects beyond Kerr paradigm}

\author{Anirban Saha}
 \altaffiliation{anirbansaha2208@gmail.com}
\affiliation{%
 School of Physical Sciences, Indian Association for the Cultivation of Science, Kolkata 700032, India\\
}%



\date{\today}

\begin{abstract}

\noindent
Multipole moments are related to the physical properties of compact gravitating objects; therefore, understanding their structure is useful in accessing the nature of compact objects. We look into gravitational wave observables for black holes with charge, black holes on the brane, black holes with torsion, and regular black holes to see if and how they are correlated to the black hole hairs, which are related to the multipole moments. We find that the gravitational wave observables are indeed related to the hairs of non-vacuum spacetimes (for instance, charge $Q$ in the case of Kerr-Newman black holes). We also constrain the black hole hairs for change in gravitational wave phasing to see if the dependencies are significant and can be observed. The results from the analysis imply that the charge $Q$ in Kerr-Newman black holes should be detectable; thus, we  provide a constraint to $Q^2/M^2$ given the spin and mass ratio of an ideal EMRI system for which future detectors like LISA can detect the change in gravitational wave observables. We also look into an analytical approach to find multipole moments of non-vacuum black hole spacetimes, mainly using the Improved Twist Vector approach for the Geroch-Hansen multipole moments and the Thorne formalism. The necessary analytics are computed, and the multipole moments are obtained for various non-vacuum spacetimes. However, the multipole moments don't contain any information about the black hole hairs, and we have commented on this observation in our paper.

\end{abstract}

\maketitle


\section{\label{sec:level1}Introduction}

        \noindent
        Multipole moments of compact objects can provide a lot of information regarding the near-horizon properties of compact objects, which can be particularly useful for distinguishing between different compact objects. The discovery of gravitational waves and the study of gravitational wave observables has opened up the study of near-horizon physics of compact objects like black holes \cite{abbott2020gw190814, abbott2020gw190425, abbott2019gwtc, abbott2017gw170817, abbott2017gw170608, abbott2016binary, abbott2016observation, PhysRevD.97.104053, Chakravarti_2020, Chakravarti_2019}. Therefore, a lot of work has been done on relating the gravitational waves of black holes with black hole hairs and, finally, relating to the analytical computation of multipole moments. Fintan D. Ryan \cite{ryan1995gravitational} demonstrated that it is possible to express the gravitational waves of compact object sources in terms of hairs and multipole moments for vacuum solutions of black holes. His work presents a method to extract multipole moments from gravitational wave observation, given certain conditions on the inspiral system that the central gravitating body is a vacuum black hole. Therefore, in this paper we intend to apply this technique\cite{ryan1995gravitational} to central object spacetimes, which are non-vacuum and by doing so, we aim to predict whether the compact objects are identifiable utilizing the gravitational wave observations to be gathered by future detectors like LISA.
        \\
        \\
        Unlike Newtonian gravitational theory, analytical analysis of multipole moments in general relativity isn't as straightforward. There are two well-known descriptions of multipole moments of compact objects. Firstly is the Geroch-Hansen formalism, which is a covariant approach to determine the multipole moments for stationary and asymptotically flat spacetimes \cite{geroch1970multipole,hansen1974multipole}. The multipolar structure is defined as a recursive function on the conformal 3-manifold of the killing vectors of the spacetime. Thomas Bäckdahl showed in his work \cite{backdahl2005explicit} that the recursive definition can be reduced to a scalar function of the norm and the twist (defined in eq. \eqref{2.1.1}) for stationary, axisymmetric and asymptotically flat spacetimes. The other approach is the one proposed by K. Thorne \cite{thorne1980multipole}. In the case of the Thorne formalism, the metric is transformed into the asymptotically Cartesian and mass-centred (ACMC) coordinates; the elements are expanded as a power series of $1/r$ and the multipole moments are extracted from the coefficients of the power series. The works  R. Geroch, Hansen \cite{geroch1970multipole,hansen1974multipole}, K. Thorne \cite{thorne1980multipole} and Fintan D. Ryan \cite{ryan1995gravitational} are done for vacuum spacetimes primarily Schwarzschild and Kerr black Holes. We intend to extend the work for non-vacuum spacetimes. Recently, there has been work to extend the existing multipole formalisms for vacuum spacetimes to non-vacuum ones. by D. Mayerson's work \cite{mayerson2023gravitational}, which we look to utilize in our paper. We shall apply the method outlined by D. Mayerson \cite{mayerson2023gravitational} and K. Thorne \cite{thorne1980multipole} to compute the multipole moments for various non-vacuum compact objects and then compare the results obtained from the different techniques.\\
        \\
       \noindent
       This paper starts with a brief summary of the spacetime we will use throughout our work in Sec. \ref{sec:level1.1}. Following this summary, in Sec \ref{sec:level2}, we shall evaluate the gravitational wave energy spectrum and gravitational wave phase evolution as the power series of $v$ (post Newtonian expansion) to work out the dependence of gravitational wave observables on the black hole hairs. Additionally, we constrain the black hole hair to see whether gravitational wave observation from future detectors can be used to infer the nature of the black holes. In Sec. \ref{sec:level3}, we discuss and analytically evaluate the multipole moments using the Improved twist approach for the Geroch-Hansen and the Thorne formalism. Finally, we discuss our results in the last Sec. \ref{sec:level4}.
       \\
        \noindent
        \textit{Notations and conventions}: In this paper, we have used $c = 1 = G$ and the positive signature convention, i.e., $\eta_{\mu\nu} = diag(-1, 1, 1, 1)$.

\section{\label{sec:level1.1}Brief Summary of the Spacetime Geometries used}

        \noindent
        In this section, we shall talk about the well-known non-vacuum and higher-dimensional geometries that we have used in this paper. The following are the space-time geometries on which we work. \\
        \\
        \noindent
        \textbf{Spacetimes with charge}: Firstly, we take a look at the  static and spherically symmetric black hole with electromagnetic charge matter, i.e. \textbf{Reissner-Nordström (RN)}  black hole with the following line element: 
             \begin{eqnarray}\label{1.1.0}
                 ds^2 = &-& \left(1 - \frac{2M}{r} + \frac{Q^2}{r^2}\right)dt^2 + \left(1 - \frac{2M}{r} + \frac{Q^2}{r^2}\right)^{-1}dr^2 \nonumber \\
                 &+& r^2d\theta^2 + r^2 \sin^2{\theta}d\phi^2
             \end{eqnarray}
        here $Q$ is the electromagnetic charge.
        \\
        We also work on the stationary, axisymmetric counterpart of RN black hole, i.e. the \textbf{Kerr-Newman (KN} black hole. The charged rotating KN black hole's line element in Boyer-Lindquist coordinates is given as follows: 
            \begin{eqnarray}\label{1.1.4}
                ds^2  = &-& \left( \frac{ \Delta - a^2 \sin^2{\theta}}{\Sigma}\right)dt^2 + \frac{\Sigma}{\Delta} dr^2 + \Sigma d\theta^2  \nonumber \\
                &+& \left( r^2 + a^2 \frac{2 M r - Q^2}{\Sigma}a^2 \sin^2{\theta}\right)\sin^2{\theta} d\phi^2  \nonumber \\
                &-& 2a\sin^2{\theta} \left( \frac{2Mr -Q^2}{\Sigma}\right) dt d\phi
            \end{eqnarray}
             \noindent here $\Delta = r^2 -2Mr + a^2 + Q^2$,$\Sigma = r^2 + a^2 \cos^2{\theta}$ where $Q$ and $a$ are charge and spin of the spacetime respectively.
        \noindent
        \\
        \\
        \textbf{Spacetime on the brane}: The gravitational field equations on the brane are determined by projecting the gravitational field equations of the higher-dimensional bulk on the brane, solving which we get the spacetimes on the brane. The static and spherically symmetric solution on the brane gives rise to the spacetime of the Braneworld black hole \cite{dadhich2000black}. The line element of the \textbf{Braneworld black holes} is:
             \begin{eqnarray}\label{1.1.1}
                 ds^2 = &-&\left(1 - \frac{2M}{r} - \frac{Q}{r^2}\right)dt^2 + \left(1 - \frac{2M}{r} - \frac{Q}{r^2}\right)^{-1}dr^2 \nonumber \\
                 &+& r^2d\theta^2 + r^2 \sin^2{\theta}d\phi^2
             \end{eqnarray}
        The structure of the metric is similar to that of the Reissner-Nordström black hole; however, the $(1/r^2)$ term is negative when $Q$ (defined as tidal charge) is positive, which isn't the case for RN black holes. Interestingly, RN black holes can be interpreted as braneworld black holes where the tidal charge is negative, leading to the  $(1/r^2)$ term being positive.
        \\
        \\

        \noindent \textbf{Spacetime with torsion}: The Kalb–Ramond field is a 2nd-rank anti-symmetric tensor which contains the signature of spacetime with torsion. Hence, the black hole spacetime with the Kalb–Ramond field, i.e.\textbf{Kalb–Ramond (KR)} black hole is the static, spherically symmetric black hole for spacetime with torsion \cite{SenGupta_2001} \cite{Kar_2003}. The metric is an infinite series, but here, we consider the metric terms up to $(1/r^3)$ order. Therefore, the line metric is as follows:
             \begin{eqnarray}\label{1.1.2}
                ds^2 = &-&\left(1 + \frac{c_1}{r} + \frac{bc_1^2}{6r^3}\right)dt^2 + \left(1 + \frac{2M}{r} - \frac{b}{r^2}\right)^{-1}dr^2 \nonumber \\
                &+& r^2d\theta^2 + r^2 \sin^2{\theta}d\phi^2
            \end{eqnarray}
        here $b$ is the parameter that represents the KR field and $c_1 = 2M$. It is evident that when $K=0$, i.e., there is no KR field, we get the Schwarzschild solution.
        \\
        \\
        \textbf{Regular Spacetimes} \noindent The \textbf{Regular} black hole proposed by Simpson and Visser is a static and spherically symmetric metric that guarantees the regularization of the singularity at $r=0$ \cite{Simpson_2019}. The line element is as follows:
              \begin{eqnarray}\label{1.1.3}
                 ds^2 = &-&\left(1 - \frac{2M}{\sqrt{r^2 + l^2}}\right)dt^2 + \left(1 - \frac{2M}{\sqrt{r^2 + l^2}} \right)^{-1}dr^2 \nonumber \\
                 &+& r^2d\theta^2 + r^2 \sin^2{\theta}d\phi^2
             \end{eqnarray}
        here, $l$ is a real and positive parameter that determines the character of regular spacetime. $l<2M$ describes the regular black hole, $l=2M$ describes a one-way wormhole and the $l>2M$ is a two-way wormhole.
        \\
        \noindent We are also interested in taking a look at regular black holes with rotation. Applying the Newman-Janis algorithm to the Simpson-Visser regular black hole metric yields the \textbf{Rotating Regular} black hole. The line element is described as follows:
            \begin{eqnarray}\label{1.1.5}
                ds^2  = &-& \left( \frac{\Delta - a^2 \sin^2{\theta}}{\Sigma}\right)dt^2 + \frac{\Sigma}{\Delta} dr^2 + \Sigma d\theta^2 \nonumber \\ 
                &+& \left( r^2 +l^2 + a^2 +  \frac{2 M \sqrt{r^2 + l^2}}{\Sigma}a^2 \sin^2{\theta}\right)\sin^2{\theta} d\phi^2 \nonumber \\
                &-& 2a\sin^2{\theta} \left( \frac{2M\sqrt{r^2 + l^2}}{\Sigma}\right) dt d\phi
            \end{eqnarray}
            \noindent
            here, $\Delta = r^2 + l^2 -2M\sqrt{r^2 + l^2} + a^2$, $\Sigma = r^2 + l^2 + a^2 \cos^2{\theta}$ where $a$ and $l$ are the spin and the real positive regularizing parameter of the spacetime. We shall work on these spacetime geometries from now onward.

\section{\label{sec:level2} Gravitational Wave Observables and associated Multipole Moments}

                \noindent
                In this section, we delve into post newtonian (PN) expansion gravitational wave observables and evaluate if future detectors like LISA could detect the observables.
                \\
                \\
                In the early inspiral phase, we consider an Extreme Mass Ratio Inspiral (EMRI) system where the larger object is central, and the orbit of the smaller compact object is geodesic, circular, equatorial, and evolves adiabatically. The central compact object is vacuum or non-vacuum, stationary, axisymmetric, reflection symmetric across the equatorial plane, and asymptotically flat. Also, we do not consider tidal heating; therefore, the lost orbital energy is radiated away as outgoing gravitational waves.
                \\
                Hence, the gravitational wave energy spectrum, i.e.the energy radiated away by the wave as the frequency evolves from $f$ to $f + df$, is equivalent to the energy $-dE$ lost from the orbit as it shrinks becomes
                   \begin{equation}\label{2.3.1}
                   \Delta E_{GW} = f\frac{dE_{wave}}{df} = -\Omega\frac{dE}{d\Omega}
                   \end{equation}

                \noindent
                The orbiting energy of the inspiraling compact object (given the above conditions) is related to the metric components and angular frequency as follows (see Appendix \ref{app2})
                    \begin{eqnarray}\label{2.3.2}
                    \frac{E}{\mu} = \frac{-g_{tt} - g_{t\phi}\Omega}{\sqrt{-g_{tt} - 2 g_{t\phi} - g_{\phi\phi}}} 
                    \end{eqnarray}
                
                \noindent
                We aim to express the gravitational wave energy spectrum in a PN expansion, which allows us to determine the central body's multipole moments $M_n$ and $S_n$ \cite{ryan1995gravitational} by analysing the coefficients at each PN order. To perform the PN expansion of the gravitational wave energy spectrum $\Delta E_{GW}$ we need to expand it in terms of the orbital frequency $\Omega$ and replace it with the linear velocity in the Newtonian limit $v$ using the relation $v = (\pi M f)^{1/3} = (M \Omega)^{1/3}$.
                \\
                \\
                \noindent
                Another gravitational wave observable we shall examine is the gravitational wave phase evolution. Since the time evolution of the gravitational wave phase can be measured with high precision, it is a valuable tool for studying gravitational waves. The system we are using is the same as we discussed earlier. So, we approach to evaluate the phase change by using
                    \begin{equation}\label{2.3.3}
                    \dfrac{d\Phi}{dt} = 2\pi f = 2 \Omega
                    \end{equation}
                Using eq. \eqref{2.3.3} and eq. \eqref{2.3.1} we can determine the phase $\Phi(t)$ change of the gravitational wave as			
    			\begin{eqnarray}\label{2.3.4}
    			d\Phi = 2 \pi f dt = 2 \pi f \frac{dE}{dE/dt}\nonumber
    			\\  
                    \Delta \Phi = 2 \pi \int_{v_i}^{v_f} \frac{\Delta E}{-dE/dt} df
    			\end{eqnarray}
                
                \noindent
                Therefore, the change in number of cycles due to the phase evolution of the wave  ($\Delta N = \frac{\Delta \Phi}{2 \pi}$) is given by			
    			\begin{equation}\label{2.3.6}
    			\Delta N =  \int_{v_i}^{v_f} \frac{\Delta E}{dE/dt} df
    			\end{equation}
                here $\frac{dE}{dt}$ is the loss of orbital energy and $\Delta E$ is the wave energy spectrum defined in eq. \eqref{2.3.2}.
                \\
                Evident from eq. \eqref{2.3.6}, we need to determine $dE/dt$ first to measure the change in the number of cycles.  Therefore, we will explore how to calculate the orbital energy loss of the inspiraling object using the method outlined in \cite{ryan1995gravitational}.
                \\
                \\
                The primary contribution of the loss of orbital energy emerges from the mass quadrupole radiative moment as shown in \cite{ryan1995gravitational}. For a given mass of central object $M$, mass of inspiral object $m$, orbital radius $\tilde{R}$ and angular velocity $\Omega$
             	  \begin{equation}\label{2.3.7}
             	-\left.\frac{d E}{d t}\right|_{I_{i j}}=\frac{32}{5} m^{2} \tilde{R}^{4} \Omega^{6}
             	\end{equation}
                \noindent
                There is also another contribution to the radiated power in from the spin moment \cite{ryan1995gravitational}
                    \begin{equation}\label{2.3.8}
                    -\left.\frac{d E}{d t}\right|_{J_{i j}}= \frac{32}{5}\left(\frac{m}{M}\right)^{2} v^{10} \left[\frac{v^{2}}{36} -\frac{1}{12} \frac{S_{v^{3}}}{M^{2}} +\frac{v^{4}}{16} \frac{S_{1}^{2}}{M^{4}} \right]
                    \end{equation}
                
                \noindent
                By summing these contributions, we determine $dE/dt$ and ultimately evaluate the change in the number of cycles.


            \subsection{\label{sec:level2.1}Gravitational wave observables in terms of black hole hairs}

                \noindent
                Now, we look at the results obtained from doing the PN expansion of gravitaional wave observables. For the sake of being organised, We shall discuss the results we found from the PN expansion of the gravitational wave energy spectrum first and then discuss the results from gravitational wave phase evolution.\\
    
                \noindent
                The gravitational wave spectrum PN expansion for Reissner-Nordstrom (RN), Braneworld (BR), Regular (Reg), Kalb-Rammond (KR) black hole are as follows:\\
                    \begin{eqnarray}\label{EGW RN}
                    \Delta E_{\mathrm{RN}} & = & \frac{v^{2}}{3}+\left(-\frac{1}{2}+ \frac{2 Q^2}{9 M^2}\right)v^4 -\left(\frac{27}{8}-\frac{ 3 Q^2}{2 M^{2}} \right.
                    \\
                    & - & \left. \frac{Q^4}{9 M^4}\right) v^{6}+\mathcal{O}\left(v^{7}\right)  \nonumber
                    \\ \nonumber
                    \\
                    \Delta E_{\mathrm{BR}} & = & \frac{v^{2}}{3}+\left(-\frac{1}{2}- \frac{2 Q}{9 M^2}\right)v^4 -\left(\frac{27}{8}+\frac{ 3 Q}{2 M^{2}} \right.
                    \\
                    & - & \left. \frac{Q^2}{9 M^4}\right) v^{6}+\mathcal{O}\left(v^{7}\right)  \nonumber
                    \end{eqnarray}
                    \begin{eqnarray}
                    \Delta E_{\mathrm{Reg}}  & = & \frac{v^{2}}{3} - \frac{v^{4}}{2}  -\frac{27}{8} v^{6} - \frac{225}{16} v^8 - \frac{6615}{128} v^{10} 
                    \\
                    & - & \frac{45927}{256} v^{12} - \left( \frac{617463}{1024} + \frac{7 l^2}{48 M^2} \right)v^{14} + \mathcal{O}\left(v^{16}\right) \nonumber
                    \\ \nonumber
                    \\
                    \Delta E_{\mathrm{KR}}  & = & \frac{v^{2}}{3}-\left(\frac{v^{4}}{2}\right)  +\left(-\frac{27}{8}+\frac{4 b}{6 M}\right) v^{6} - \left(\frac{225}{16} \right. 
                    \\
                    & - & \left. \frac{20 b}{6 }\right) v^8 +\mathcal{O}\left(v^{9}\right) \nonumber
                    \end{eqnarray}
                
                \noindent
                Next are the results for PN Expansion for gravitational wave energy spectrum for rotating non-vacuum spacetimes Kerr-Newman (KN) and Rotating Regular (RR) black hole.
                    \begin{eqnarray}\label{EGW KN}
                    \Delta E_{\mathrm{KN}} & = & \frac{v^{2}}{3}+\left(-\frac{1}{2}+ \frac{2 Q^2}{9 M^2}\right)v^4 + \frac{20a}{9M} v^5 
                    \\
                    & - & \left( \frac{27}{8}  + \frac{a^2}{M^2} - \frac{ 3 Q^2}{2 M^{2}} -  \frac{Q^4}{9 M^4} \right) v^{6} \nonumber
                    \\
                    & + & \frac{28}{3}\left(\frac{a}{M} -  \frac{a Q^2}{9 M^2}\right) v^7 + \mathcal{O}\left(v^{8}\right) \nonumber
                    \\
                    \Delta E_{\mathrm{RR}} & = & \frac{v^{2}}{3} - \frac{v^{4}}{2}  +\frac{20a }{9 M} v^{5}  -\left( \frac{27}{8} + \frac{a^{2}}{M^{2}} \right) v^{6} 
                    \\
                    & - & \left( \frac{225}{16} + \frac{130}{27} \frac{a^2}{M^2}\right) + \frac{81}{2} a v^9  - \left( \frac{6615}{128} \right. \nonumber
                    \\
                    & + &  \left. \frac{2345}{72} \frac{a^2}{M^2}\right) v^{10}  + \left( \frac{242}{27} \frac{a^3}{M^3} + 165 \frac{a}{M} \right) v^{11} \nonumber
                    \\
                    & - & \left( \frac{45927}{256} + \frac{3}{2} \frac{a^4}{M^4} + \frac{5}{9} \frac{a^2 l^2}{M^4} + \frac{1377}{8} \frac{a^2}{M^2} \right) v^{12} \nonumber
                    \\
                    & - & \left( \frac{13}{24} \frac{a}{M^4} - \frac{1976}{27} \frac{a^3}{M^3} - \frac{20475}{32} \frac{a}{M} \right)v^{13} \nonumber
                    \\
                    & - & \left( \frac{617463}{1024} + \frac{104335}{128} \frac{a^2}{M^2} + \frac{26831}{1944} \frac{a^4}{M^4}   \right. \nonumber
                    \\
                    & - & \left. \frac{35}{36} \frac{a^2 l^2}{M^4} + \frac{7 l^2}{48 M^2}  \right)v^{14} + \mathcal{O}\left( v^{16}\right) \nonumber
                    \end{eqnarray}
                
                \noindent
                The gravitational wave energy spectrum expressions contain contributions from $Q^2$ for the RN black hole, $l^2$ for the Regular black hole, $b$ for the KR black hole, $Q^2$ and $a$ for the KN black hole and $l^2$ and $a$ for Rotating Regular black hole.
                
                \noindent
                Given the differences in the gravitational wave energy spectrum between non-vacuum and vacuum solutions, we shall evaluate the gravitational wave phase evolution for rotating non-vacuum black holes as well and compare it to that of the Kerr solution to find the differences in phase evolution.
                    \begin{eqnarray}\label{no of cycles for KN}
                    \Delta N_{KN} & = &  \frac{- 5 M}{32 \pi m} v^{-5}  \left[ \frac{1}{5} + \left( \frac{743}{1008} + \frac{2 Q^2}{9 M^2}\right) v^2 \right.
                    \\
                    & - & \left( 2 \pi + \frac{113}{24} \frac{a}{M}  \right) v^3 + \left( \frac{3058673}{1016064}  - \frac{81}{16} \frac{a^2}{M^2} \right. \nonumber
                    \\
                    & + & \left. \left. \frac{3515}{504} \frac{Q^2}{M^2} + \frac{1}{3} \frac{Q^4}{M^4} \right) v^4 +  \mathcal{O}(v^5 ) \right] \nonumber
                    \end{eqnarray}
                    \begin{eqnarray}
                    \Delta N_{RR} & = &  \frac{- 5 M}{32 \pi m} v^{-5}  \left[ \frac{1}{5} + \frac{743}{1008}v^2 - \left( 2 \pi + \frac{113}{24} \frac{a}{M} \right) v^3 \right. \nonumber
                    \\
                    & + & \left. \left( \frac{3058673}{1016064} - \frac{2 l^2}{M^2} - \frac{81}{16} \frac{a^2}{M^2} \right) v^4 + \mathcal{O}(v^5 )\right] 
                    \end{eqnarray}
                    
                \noindent
                The phase evolution and the energy spectrum of gravitational waves are related to the non-vacuum hairs of each black hole. For instance, in the case of Kerr-Newman black holes, the gravitational wave observables depend on $Q^2/M^2$ and $a/M$, and for Rotating regular black holes, they are related with $l^2/M^2$ and $a/M$. Therefore, the gravitational wave's phase evolution and energy spectrum for non-vacuum and vacuum systems differ. However, there are notable differences in the PN expansions for the Kerr-Newman and Rotating Regular black holes. Specifically, charge $Q^2$ appears at a lower PN order $(PN-2)$ than the angular momentum $a$ at $PN-3$. Whereas, for the Rotating Regular black hole parameter $l^2$ appears at higher PN order $(PN-4)$ compared to angular momentum at $PN-3$.
                \\
               \subsection{\label{sec:level2.2}Constraints on the hairs of rotating non-vacuum black holes}
                \noindent
                We wish to constrain the value of $Q^2/M^2$ and $l^2/M^2$ to see whether measurable changes in phase evolution of gravitational waves can be produced with plausible constraints on $Q^2/M^2$ and $l^2/M^2$. To evaluate the difference in phase evolution of gravitational waves, we define the fractional change in gravitational wave phase evolution as follows:
                    \begin{eqnarray}
                    \Delta N_{frac} =  \frac{\Delta N_{non-vac} - \Delta N_{vac}}{\Delta N_{vac}}
                    \end{eqnarray}
    
                \noindent
                In the above equation, we shall evaluate the change in cycles for Kerr-Newman and Rotating Regular black holes and take the Kerr vac solution to compare it. The results for constraining $l^2/M^2$ and $Q^2/M^2$ for a measurable change in the gravitational wave phase evolution are as follows.
                    \begin{table}[htbp]
                    \caption{\label{tab:table 4.4} Constraining $l^2/M^2$ for Rotating Regular hole with $\Delta N_{frac} = 0.1$ , $m = 10^2 M_{\odot} $ , $v_i = 0.15$ and $v_f = 0.25$}
                    \begin{center}
                    \begin{tabular}{  P{4em}  P{4cm} P{3cm}  } 
                     \hline
                    \\
                    $M/m $ & $a/M$ & $l^2/M^2$ \\
                    \\
                    \hline
                    \hline
                    \\
                    & 0 & 43.50\\
                    & 0.20 & 43.39\\	
                    $10^2$  & 0.30 & 43.34\\ 			
                    & 0.40 & 43.27\\
                    \\
                    \hline
                    \\
                    & 0 & 43.50\\
                    & 0.20 & 43.39\\ 			
                    $10^3$ & 0.30 & 43.34\\ 			
                    & 0.40 & 43.27\\
                    \\
                    \hline
                    \\
                    & 0 & 43.50\\
                    & 0.20 & 43.39\\ 			
                    $10^4$ & 0.30 & 43.34\\ 			
                    & 0.40 & 43.27\\
                    \\
                    \hline
                    \end{tabular}
                    
                    \end{center}
                    \end{table}
    
                \noindent
    
                \noindent
                The results from constraining $l^2/M^2$ for the Rotating Regular hole in table \ref{tab:table 4.4} indicate that for a measurable difference of change in gravitational wave phase from Kerr black hole, we consistently find $l^2/M^2 \gg 1$ which isn't realistic. This can somewhat be anticipated given that the contribution of $l^2/M^2$ is at a higher ON order than $a/M$. Unfortunately, it implies that distinguishing between Kerr and Rotating regular Black holes isn't possible without higher precision.
                    \begin{table}[htbp]
                    \caption{\label{tab:table 4.5} Constraining $Q^2/M^2$ for Kerr-Newman black hole with $\Delta N_{frac} = 0.1$ , $m = 10^2 M_{\odot} $ ,  $v_i = 0.303$ and $v_f = 0.377$}
                    \begin{center}
                    \begin{tabular}{ P{4em} P{4cm} P{3cm}} 
                    \hline
                    \\
                    $M/m $ & $a/M$ & $Q^2/M^2$ \\
                    \\
                    \hline
                    \hline
                    \\
                    & 0 & 0.213\\
                    & 0.20 & 0.254\\	
                    $10^2$  & 0.30 & 0.272\\ 			
                    & 0.40 & 0.289\\
                    \\
                    \hline
                    \\
                    & 0 & 0.213\\
                    & 0.20 & 0.254\\ 			
                    $10^3$ & 0.30 & 0.272\\ 			
                    & 0.40 & 0.289\\
                    \\
                    \hline
                    \\
                    & 0 & 0.213\\ 
                    & 0.20 & 0.254\\ 			
                    $10^4$ & 0.30 & 0.272\\ 			
                    & 0.40 & 0.289\\
                    \\
                    \hline
                    \end{tabular}
                    
                    \end{center}
                    \end{table}
                    
                \noindent
                In Table \ref{tab:table 4.5}, we find a measurable difference in the evolution gravitational wave phase from the Kerr black hole with the $Q^2/M^2 < 1$. To improve the constraint on $Q^2/M^2$, we used $v_i = 0.303$ and $v_f = 0.377$ for the initial and final velocities of the inspiraling object, respectively. The final velocity corresponds to its velocity at the ISCO (innermost stable circular orbit), and its initial velocity is the velocity of the orbiting object four years before reaching the ISCO. The result indicates that if Kerr-Newman has enough charge, it should be distinguishable from Kerr black holes through observations from future detectors like LISA.
            
            \section{\label{sec:level3}Analytical Approach to Multipole Moments}

            \noindent
            In vacuum spacetimes, the theoretical multipole moments of a compact object can be used to evaluate gravitational wave observables, and conversely, determining multipole moments from gravitational wave observables should yield consistent results. This is because the multipole moments from gravitational wave data and the calculated multipole moments agree with each other. 
                    \\
            Now, we shall explore the same for several non-vacuum spacetimes, including the Reissner-Nordström, Kalb-Ramond, Braneworld, and Regular black holes. For rotating non-vacuum black hole solutions, we examine the Kerr-Newman and Rotating Regular black holes.

            
            \subsection{\label{sec:level3.1} Evaluating Multipole Moments of non-vacuum black hole solution using improved twist approach of Geroch-Hansen multipoles}

                \noindent In this section, we discuss one of the methods to evaluate multipole moments of non-vacuum compact objects. The Improved twist approach of Geroch-Hansen multipoles modifies the twist vector in the Geroch-Hansen multipole formalism for vacuum, axisymmetric, stationary and asymptotically flat spacetimes \cite{geroch1970multipole,hansen1974multipole}. First, we define the original norm and twist vector as 
                    \begin{eqnarray}\label{2.1.1}
                         &\lambda  = -\xi^a\xi_a \nonumber \\
                         &\omega_\mu = -\epsilon_{\mu\nu\rho\delta}\xi^\nu \nabla^\rho\xi^\delta
                     \end{eqnarray}
                     
                \noindent 
                It can be shown that $\nabla_{[\mu}\omega_{\nu]} = -\epsilon_{\mu\nu\rho\delta}\xi^\rho R^\delta_\lambda\xi^\lambda$. Therefore, the curl of the twist is non-zero for non-vacuum spacetimes, which does not allow us to define a twist scalar whose gradient gives the twist vector. To circumvent this problem \cite{mayerson2023gravitational} shows that one can define an additional twist vector called the improved twist vector, which satisfies:
                    \begin{eqnarray}\label{2.1.1.0}
                         \nabla_{[\mu}\omega_{\nu]}^I = \epsilon_{\mu\nu\rho\delta} \xi^\rho T^\delta_\lambda \xi^\lambda
                    \end{eqnarray}
                     
                \noindent
                From eq. \eqref{2.1.1.0} we define $\omega^{total}_{\nu} = \omega_{\nu} + \omega^{I}_{\nu}$ and we can see it satisfies $\nabla_{[\mu}\omega^{total}_{\nu]} = 0$,  for vacuum and non-Vacuum spacetimes. Using this we shall determine the multipole moments of non-vacuum spacetimes.\\

                \noindent
                Now, we follow the same process of evaluating Geroch-Hansen multipole moments using the $\omega^{total}$ instead of $\omega$.
                Therefore, the induced metric $\tilde h_{ab} = \Omega^2(\Lambda g_{ab} + \xi_a\xi_b)$ which in stationary, axis-symmetric  and asymptotically flat spacetimes should take the form:
                    \begin{eqnarray}\label{2.1.2}
                        ds^2 =  dR^2 + R^2d\theta^2 +  e^{-2\beta}R^2 \sin^2{\theta}d\phi^2
                    \end{eqnarray}

                \noindent
                Now, the Ernst potential is defined in terms of the norm ($\lambda$) and total twist (abbreviating $\omega^{total}$ to $\omega^t$) as,
                    \begin{eqnarray}\label{2.1.3}
                        \phi = \frac{\lambda^2 + \omega^{{t}^2} - 1 + iw^{t}}{4\lambda} 
                    \end{eqnarray}
                    
                \noindent
                It is then conformally transformed $\phi_H \rightarrow \tilde \phi_H = \phi_H/\sqrt{\Omega} $. Finally, we make the transformation $\rho = R\sin{\theta}$ and $z = R\cos{\theta}$. Finally put $\rho = -iR$, $z = R$ to get to the leading order function (see Definition 4 in \cite{backdahl2005explicit}).\\
                    
                \noindent
                Next, we  require a few new functions related to the metric to evaluate the multipole moments: $y = e^{k_L} \phi_{L}$ where $\rho = R e^{k_{L} -\beta_{L}}$ and $k_L = -\ln{(1 - R \int_0^R \frac{\left(e^{2 \beta_{L}} -1\right)}{R^2} dR)} + \beta_L$ and $\beta_L = \beta(R,i R)$ (from eq. \eqref{2.1.2}). Finally, the final multipole moments are expressed as:
                    \begin{eqnarray} \label{2.1.4}
                        &M_n = \left. \frac{2^n n!}{(2 L)!}\frac{\partial^n y}{\partial \rho^n}\right|_{\rho = 0}
                    \end{eqnarray}

            \subsection{\label{sec:level3.2} Evaluating Multipole Moments of non-vacuum black hole solution using Thorne multipole formalism}

                \noindent
                We shall also work with the formalism developed by Thorne in \cite{thorne1980multipole}. This requires the "Asymptotically Cartesian and Mass-Centered" (ACMC) coordinates for asymptotically flat, stationary, and vacuum spacetimes.
                \\
                There are primarily two conditions for a coordinate system to qualify as ACMC. The first is being asymptotically flat. A coordinate system is asymptotically Cartesian when the metric in those coordinates, when expanded, contains terms at order $r^{-(\ell+1)}$ which have angular dependence on spherical harmonics till order $\ell$. A coordinate system with a higher-order angular dependence than this can't be ACMC. Now, the second condition for an ACMC coordinate is the mass centered, which implies the $r^{-2}$ term in $g_{00}$ is constant (no mass dipole). Once the metric is in ACMC coordinates, we can extract the multipole moments by comparing the expansion of the metric elements with the following equations.
                \\
                    \begin{align}     
                    & g_{t t} = - 1+\frac{2 M}{r} +  \sum_{a \geq 1}^{n} \frac{2}{r^{a+1}}\left({M}_{a} P_{a} + \sum_{b < a} c_{a b}^{(t t)} P_{b}\right) \label{2.2.1.a}
                    \\
                     &+ \frac{2}{r^{n+2}}\left({M}_{n+1} P_{n+1} + \sum_{b \neq n+1} c_{(n+1) b}^{(t t)} P_{b}\right) + \mathcal{O}\left(\frac{1}{r^{(n+3)}}\right) \nonumber
                    \\
                    & g_{t \phi}    = - 2 r\sin^2 \theta\left[\sum_{a \geq 1}^{n} \frac{1}{r^{a+1}}\left(\frac{{S}_{a}}{a} P_{a}^{\prime}+\sum_{b<a} c_{a b}^{(t \phi)} P_{b}^{\prime}\right)\right. \label{2.2.1.b}
                    \\
                     &\left. + \frac{1}{r^{n+2}}\left(\frac{{S}_{n+1}}{n+1} P_{n+1}^{\prime}+\sum_{b\neq n+1} c_{(n+1) b}^{(t \phi)} P_{b}\right) + \mathcal{O}\left(\frac{1}{r^{(n+3)}}\right)\right] \nonumber
                    \\
                    & g_{r r}   = 1 + \sum_{a \geq 0}^{n} \frac{1}{r^{a+1}} \sum_{b \leq a} c_{a b}^{(r r)} P_{a}^{\prime} +  \frac{1}{r^{n+2}} \sum_{b} c_{(n+1) b}^{(r r)} P_{a} \nonumber 
                    \\
                    & + \mathcal{O}\left(\frac{1}{r^{(n+3)}}\right) \label{2.2.1.c}
                    \end{align}
                    
                    \begin{align}
                    g_{\theta \theta}  = r^{2} &\left[1+\sum_{a \geq 0}^{n} \frac{1}{r^{a+1}} \sum_{b \leq a} c_{a b}^{(\theta \theta)} P_{a} \right.\label{2.2.1.d}
                    \\
                    &\left. + \frac{1}{r^{n+2}} \sum_{b } c_{(n+1) b}^{(\theta \theta)} P_{a}+\mathcal{O}\left(\frac{1}{r^{(n+3)}}\right)\right] \nonumber
                    \\
                    g_{\phi \phi}  = r^{2} & \sin ^{2} \theta\left[1+\sum_{a \geq 0}^{n} \frac{1}{r^{a+1}} \sum_{b \leq a} c_{a b}^{(\phi \phi)} P_{a} \right. \label{2.2.1.e}
                    \\
                    & \left. + \frac{1}{r^{n+2}} \sum_{b } c_{(n+1) b}^{(\phi \phi)} P_{a} + +\mathcal{O}\left(\frac{1}{r^{(n+3)}}\right)\right] \nonumber
                    \\
                    g_{r \theta}  = -r &\sin \theta \left[\sum_{a \geq 0}^{n} \frac{1}{r^{a+1}} \sum_{b \leq a} c_{a b}^{(r \theta)} P_{a}^{\prime} \right.\label{2.2.1.f} 
                    \\
                    &\left. + \frac{1}{r^{n+2}} \sum_{b } c_{(n+1)) b}^{(r \theta)} P_{a}^{\prime}+\mathcal{O}\left(\frac{1}{r^{(n+3)}}\right)\right] \nonumber
                    \end{align}

                \noindent
                Here, $r$ represents the radial coordinate, $P_l$ is a Legendre polynomial with $\cos{\theta}$ as its argument, and $c_{ab}$ are constants. The terms $M_a$ correspond to mass moments and $S_a$  to spin multipole moments, and as is evident, $M_a$ and $S_a$ can be determined derived from the components $g_{tt}$ and $g_{t\phi}$ respectively. 
                \\
                We shall, therefore, use this Throne formalism to evaluate the multipole moments of non-vacuum spacetimes mentioned before.

            \subsection{Results from improved twist approach of Geroch-Hansen multipoles and Thorne formalism }
            
            \noindent
            Next, we see the results we found for several non-vacuum, axisymmetric, stationary and asymptotically flat spacetimes, i.e. Reissner-Nordstrom, Kerr-Newman Kalb-Ramond, Braneworld, Regular and Rotating Regular black hole using the improved Twist approach of Geroch-Hansen multipole or the Throne multipole moment formalism.\\
            
            \noindent
            Note that the Regular black holes' multipole moments are categorised into:  Case I: $l \gg 2M$ and Case II: $ l \ll 2M$. These assumptions are necessary because, without them, the coordinate transformation required to convert the induced metric $\tilde{h}_{ab}$ in its required form in eq. \eqref{2.1.2} isn't invertible (Appendix \ref{regular} for detailed calculation).
            \\
            
            \noindent
            First, we let us look at the results from non-rotating spacetimes, then move on to the rotating non-vacuum solutions. The results for non-rotating solutions can be found in Table \ref{tab:table 4.1}  and \ref{tab:table 4.2},
            
                \begin{table}[htbp]\caption{\label{tab:table 4.1} Multipole moment for non-vacuum, stationary, non-rotating, asymptotically flat black hole spacetimes evaluated using the Improved Twist Approach of the Geroch and Hansen formalism (Appendix \ref{regular1} for detailed calculations)}
                \begin{center}
                \begin{tabular}{  P{2em}  P{3cm} P{4.5cm}  } 
                \hline
                \\
                Sl no.& Black Hole Spacetimes &  Multipole Moments using Method \ref{sec:level3.1} \\
                \\\hline
                \\
                1 & Reissner–Nordstrom & $M_0 = -M, M_n = 0$ \\
                2 & Kalb-Ramond & $M_0 = -M, M_n = 0$  \\
                3 & Braneworld & $M_0 = -M, M_n = 0$  \\
                4 & Regular \footnotemark[1]  & Case I : $M_0 = -M, M_n = 0$ \\
                & & Case II : $M_0 = -M, M_n = 0$\\
                \\
                \hline
                \end{tabular}
                \footnotetext[1]{the solution for the Regular Black hole is for two cases,
                \\
                Case I: $l \gg 2M$ and Case II: $ l \ll 2M$}
                \end{center}
                \end{table}

                \begin{table}[htbp]\caption{\label{tab:table 4.2} Multipole moment for non-vacuum, stationary, non-rotating, asymptotically flat black hole spacetimes evaluated using the Throne formalism (Appendix \ref{regular2} for detailed calculations)}
                \begin{center}
                \begin{tabular}{  P{2em}  P{3cm} P{4.5cm}  } 
                \hline
                \\
                Sl no.& Black Hole Spacetimes &  Multipole Moments using Method \ref{sec:level3.2} \\
                \\\hline
                \\
                1 & Reissner–Nordstrom & $M_0 = -M, M_n = 0$ \\
                2 & Kalb-Ramond & $M_0 = -M, M_n = 0$  \\
                3 & Braneworld & $M_0 = -M, M_n = 0$  \\
                4 & Regular  & $M_0 = -M, M_n = 0$ \\
                \\
                \hline
                \end{tabular}
                
                \end{center}
                \end{table}
                
            \noindent 
            It is evident that multipole moment results for all the spacetimes evaluated are consistent across both formalisms. However, interestingly, the multipole moments evaluated by both formalisms are identical to those of the Schwarzschild black hole for all the black hole spacetimes, suggesting that despite evaluating non-vacuum spacetimes, the non-vacuum hairs don't appear in the multipole moments. Now, lets take a look at the results for the rotating spacetimes to see whether we find a similar trend as well.\\

            \noindent
            Now, evaluating the multipole moment for rotating black holes, we face a problem. Unlike the non-rotating case, we don't have a trivial solution for the improved twist. Also, since we don't have a discrete definition of the improved twist, we have proceeded using the Thorne formalism \ref{sec:level3.2} only. It would at least allow us to see if there are any changes in the non-vacuum rotating cases from the multipole moments of the Kerr solution.\\

                \begin{table}[htbp]\caption{\label{tab:table 4.3} In this table are the results of the Multipole moment for non-vacuum, stationary, non-rotating, asymptotically flat black hole spacetimes (see Appendix \ref{app4} for detailed calculations)}
                \begin{center}
                \begin{tabular}{  P{2em}  P{3cm} P{4.5cm}  } 
                \hline
                \\
                 Sl no. & Black hole Sptm &  Multipole moments using Method \ref{sec:level3.2}\\
                 \\\hline
                 \\
                 1 & Kerr Newman &  $ M_n = -M (- \iota a)^n$\\
                 2 & Rotating Regular &  $ M_n = -M (- \iota a)^n $\\
                 \\
                \hline
                \end{tabular}
                
                \end{center}
                \end{table}

            \noindent
            The results show that the calculated multipole moments aren't affected by the non-vacuum properties as they are equivalent to that of the Kerr solution for both the considered spacetimes. This is a consistent trend for both the rotating and non-rotating spacetimes; however, as we know from the PN expansion, the gravitational wave observables depend on the non-vacuum characteristics, which are not present in the multipole moments evaluated using these formalisms. Thus, we are unable to express the gravitational wave observables in terms of the theoretical multipole moments for non-vacuum spacetimes.

\section{\label{sec:level4}Concluding Remarks}

\noindent
Gravitational wave quantities like gravitational wave energy spectrum and phase evolution are expected to contain the signatures of the hairs of compact objects. Evaluating for an EMRI Inspiral system (smaller object orbits a central compact in a geodesic circular and equatorial orbit and evolves adiabatically), we find that the gravitational waves observables for the non-vacuum central objects deviate significantly from the Kerr for the rotating non-vacuum central compacts objects and Schwarzschild black holes in case of spherically symmetric compact objects. It is interesting to find that charge $Q$ contributes at a lower PN order compared to the spin $a$ for the Kerr-Newman solution. In contrast, the parameter $l$ for the Rotating Regular black hole appears at a higher PN order than $a$.
We have also determined constraints for the non-vacuum black hole hairs by parameterising gravitational wave phase evolution. For an observational difference in gravitational waves phase evolution ($\Delta N_{frac} \geq 0.1$) for Kerr-Newman is possible with $Q^2/M^2 < 1$, which is plausible. However, in the case of rotating regular black holes, a significant deviation in phase evolution occurs if $l^2/M^2 \gg 1$, which isn't realistic. This is expected to some extent since the contribution of $l$ is at a higher PN order when compared to $Q$. We  find a good constraint to $Q^2/M^2$ given the spin and mass ratio of an EMRI system for which gravitational wave observable can be distinguished by future detectors like LISA. For instance, for an ideal EMRI system  with  mass ratio $M/m= 10^4$ and spin of the central gravitating object $a/m= 0.3$ we expect the differences in gravitational wave phase evolution to be detectable given $Q^2/M^2 \geq 0.272$. In general, we can conclude that the difference in gravitational wave phase evolution in the inspiral of an EMRI system with central Kerr and Kerr-Newman black hole with significant value of $Q^2/M^2$ should be observable with future detectors like LISA.
\\
\\
In the second part of our work, we have discussed an analytical approach to multipole moments for different non-vacuum spacetimes using the improved twist approach for Geroch-Hansen formalism and Throne formalism. Our analysis shows that the multipole moments derived from both formalisms are consistent with each other. The multipole moments for Reissner-Nordstrom, Braneworld, and Regular are equivalent to that of the Schwarzschild solution. Additionally, The multipoles of Kerr-Newman and rotating Regular black holes are equal to the multipoles of the Kerr solution. However, the gravitational wave observables contain contributions from the non-vacuum hairs, which couldn't be replicated from the improved twist approach for Geroch-Hansen multipole moments and the Throne formalism. We suspect an exact solution for the Improved Twist could potentially resolve this discrepancy, and addressing it is part of our future work.
\\

\begin{acknowledgments}
                
\noindent
I would like first to express my profound gratitude and deep regard to my Master's Supervisor, Dr Sumanta Chakraborty, Indian Association for the Cultivation of Science, Kolkata. I sincerely wish to acknowledge his vision, guidance, and valuable feedback. I am also grateful to the Indian Association for the Cultivation of Science, Kolkata, for providing the necessary resources and facilities to complete this work to the best of my ability.
                
\end{acknowledgments}


\appendix

        \section{Appendixes}\label{app}
                   
            \subsection{Relation for the dependence of the energy to orbiting object to the metric} \label{app2}
            
            \noindent
            In this Appendix, let us evaluate the form given in EQ. \eqref{2.3.2} . First, the standard normalisation of 4-velocity gives         
            \begin{eqnarray}
                &u^au_a = -1 \nonumber \\
                &g_{ab} u^au^b = -1 \nonumber
            \end{eqnarray}
            \begin{eqnarray}\label{90}
                -1 = g_{tt} \left(\frac{dt}{d\tau}\right)^2 + 2g_{t\phi}\left(\frac{dt}{d\tau}\frac{d\phi} {d\tau}\right) + g_{\phi\phi}\left(\frac{d\phi}{d\tau}\right)^2 \\
                + g_{rr}\left(\frac{dr}{d\tau}\right)^2 + g_{\theta\theta}\left(\frac{d\theta}{d\tau}\right)^2 \nonumber
            \end{eqnarray}
            We know that stationary, axisymmetric metric has $d/dt$ and $d/d\phi$ killing vectors; hence, the metric is not dependent on t and $\phi$. Therefore, the energy per unit mass should be conserved. To evaluate it we take $p^t$            
            \begin{equation}
                p_t = - E =  m[g_{tt} u^t + g_{t\phi} u^\phi] \nonumber
            \end{equation}
            using $u^t = \left(\frac{dt}{d\tau}\right)$ and $u^\phi = \left(\frac{d\phi}{d\tau}\right)$            
            \begin{equation}\label{92.1}
                \frac{E}{m} = - g_{tt} \left(\frac{dt}{d\tau}\right) - g_{t\phi}\left(\frac{d\phi}{d\tau}\right)
            \end{equation}
            
            \noindent
            We assumed earlier that the inspiral body follows a geodesic path and falls into the central body adiabatic. Hence, it follows the Geodesic equation.            
            \begin{equation}
                \frac{du^m}{d\tau} + {\Gamma^m}_{ab} u^a u^b = 0 \nonumber
            \end{equation}
            Taking $m = r$ we arrive at the following result:
            \begin{eqnarray}
                g_{tt,r} + 2g_{t\phi , r} \left(\frac{d\phi}{dt}\right) + g_{\phi \phi, r}\left(\frac{d\phi}{dt}\right)^2 = 0 \nonumber
            \end{eqnarray}
            Solving the quadratic equation gives us            
            \begin{equation}\label{104.1}
                \Omega = \left(\frac{d\phi}{dt}\right) = \frac{g_{t\phi,r} + \sqrt{(g_{t\phi}, r)^2 - g_{tt, r}g_{\phi\phi, r}}}{g_{\phi\phi, r}}
            \end{equation}
            From chain rule $\frac{d\phi}{d\tau} = \Omega\frac{dt}{d\tau}$. Also, for a circular orbit at the equatorial plane we have $\theta = 0$, and $r =$ constant therefore, $\frac{dr}{t} = 0$ and $\frac{d^2r}{dt^2} = 0$ the eq. \eqref{90} now becomes\\
            
            \begin{eqnarray}\label{105.1}
                -1 = g_{tt} \left(\frac{dt}{d\tau}\right)^2 + 2g_{t\phi}\left(\frac{dt}{d\tau}\frac{d\phi} {d\tau}\right) + g_{\phi\phi}\left(\frac{d\phi}{d\tau}\right)^2 \nonumber
            \end{eqnarray}
            Substituting yields
            \begin{equation}\label{106.1}
                \left(\frac{dt}{d\tau}\right) = \frac{1}{\sqrt{-g_{tt} - 2\Omega g_{t\phi} - \Omega^2g_{\phi\phi}}} 
            \end{equation}
            
            \noindent
            Now just substituting eq. \eqref{106.1} and eq. \eqref{104.1} into eq. \eqref{92.1} gives us the expression of $\frac{E}{m}$as   
            \begin{equation}\label{E/m}
                \frac{E}{m} = \frac{-g_{tt} - g_{t\phi}\Omega}{\sqrt{-g_{tt} - 2 g_{t\phi} - g_{\phi\phi}}}
            \end{equation}

            \noindent
            Using eq. \eqref{E/m} and eq. \eqref{2.3.1} we can evaluate the gravitational wave energy spectrum, using which we determine the change in number of cycles due to the gravitational wave phase evolution.

            \subsection{Multipole Moments of Regular Black Hole} \label{regular}


                    \subsubsection{ Evaluating Multipole Moments of Regular Black Holes using improved twist form of Geroch-Hansen multipoles}  \label{regular1}

                    \noindent
                    In this section, we use the Improved twist form of Geroch-Hansen multipoles to compute the multipole moments of Regular black hole \cite{Simpson_2019} which have no singularity at $r = 0$
                    \begin{eqnarray}\label{2.3.1.1}
                         ds^2 = &-&\left(1 - \frac{2M}{\sqrt{r^2 + l^2}}\right)dt^2 + \left(1 - \frac{2M}{\sqrt{r^2 + l^2}} \right)^{-1}dr^2 \nonumber \\
                         &+& r^2d\theta^2 + r^2 \sin^2{\theta}d\phi^2
                     \end{eqnarray}
                     here, $l$ $l$ is a real and positive parameter that determines the character of regular spacetime.
                     \\
                     $\xi^t$ is the timelike killing vector field. Using it we evaluate the norm and get         
                    \begin{eqnarray}
                         &\lambda  = -\xi^a\xi_a = - g_{tt}\xi^t\xi^t = \left(1 - \frac{2M}{\sqrt{r^2 + l^2}}\right)
                     \end{eqnarray}
                    Now evaluating the Improved twist we do the follows;           
                    \begin{eqnarray}
                        \partial_{[\mu}\omega^I_{\nu]} = \epsilon_{\mu\nu\rho\sigma} \xi^\rho T^\sigma_\lambda \xi^\lambda = \epsilon_{\mu\nu 0 \sigma} T^\sigma_0 \nonumber
                     \end{eqnarray}
                    $T^3_0, T^2_0, T^1_0$ are the ones that contribute, but since regular black holes have a diagonal metric, the non-diagonal elements of the Ricci tensor are zero. Therefore, we have  $T^3_0 = T^2_0 = T^1_0 = 0$. So,            
                     \begin{equation}
                         \nabla_{[\mu}\omega^I_{\nu]} = 0
                     \end{equation}
                     It is clear that $\omega^I_{\nu]}$ can be represented as $\nabla_\nu \omega^I$, which we can conveniently take as zero. Therefore, the improved twist is 
                     \begin{eqnarray}
                         &\omega^I = 0 \nonumber \\
                         &\omega^{total} = \omega + \omega^I = 0
                    \end{eqnarray}

                     
                    \noindent
                    The transformation of from $r$ to $R$  can be derived from the following form.  \begin{eqnarray}
                        \int \frac{x dx}{\sqrt{x^2 - l^2}\sqrt{x^2 - 2M x}} = \int \frac{dR}{R}
                    \end{eqnarray} 
                    where $x = \sqrt{r^2 +l^2}$.
                    Unfortunately, to find the integral is in the form of Elliptical functions and since we require $r$ in terms of $R$ we need to invert the expression. Since, elliptical functions are difficult to invert we take couple of different approximations to make our calculation simpler. They are as follows:\\

                    \noindent
                    \textbf{Case I: $l >> 2 M $}\\
                    For this case we find the transformation and conformal factor to be
                    \begin{eqnarray} \label{2.3.1.2}
                        &r = \sqrt{u^2 - l^2} = \left(\frac{1}{R} - \frac{R}{4}l^2\right) \nonumber \\
                        &\Omega = \frac{R^2}{\sqrt{\left(\left(1 + \frac{R^2}{4}l^2\right)^2 - 2 M R\left(1 + \frac{R^2}{4}l^2\right)\right)}}
                    \end{eqnarray}

                    \noindent
                    Following the necessary steps described in Sec. \ref{sec:level3.1} we arrive at the multipole moments for the Regular Black hole;
                    \begin{eqnarray}\label{2.3.1.3}
                        &M_0 = - M \nonumber \\
                        &M_1 = 0, M_2 = 0     
                    \end{eqnarray}
                    The higher-order multipole moments are also zero.
                    \\        
                    Therefore, the multipole moments of the Regular black holes in the limit $l >> 2 M$ match that of the Schwarzschild solution. Next, lets attempt to find out the multipole moment in the limit $l << 2 M$.\\

                    
                    \noindent
                    \textbf{Case II: Limit $l << 2 M$}\\
                    Given this approximation, $l << 2 M $, we can evaluate the integral by ignoring the $l$ term. The transformation finally becomes,
        
                    \begin{equation} \label{2.3.1.4}
                        r = \frac{1}{R}\sqrt{\left(1 + MR + \frac{M^2R^2}{4}\right)^2 - l^2R^2}
                    \end{equation}
                    Now, we take a look at the conformal factor,
                    \begin{eqnarray}
                        \Omega  = \frac{R}{\sqrt{r^2 + l^2 - 2 M \sqrt{r^2 + l^2}}}
                    \end{eqnarray}

                    \noindent
                    Following the necessary steps shown in Sec. \ref{sec:level3.1} we arrive at the multipole moments for the Regular Black hole;
                    \begin{eqnarray}\label{2.3.1.5}
                        &M_0 = - M  \\
                        &M_1 = 0, M_2 = 0  \nonumber   
                    \end{eqnarray}
                    The higher-order multipole moments are also zero.
                    \\        
                    Hence, it is evident that the multipole moment derived from the limit of a regular black hole in \textbf{Case II} matches with that in \textbf{Case I}, which is the same as that of the Schwarzschild solution. Next, we can attempt to evaluate the multipole moments using the Thorne formalism and check whether the results agree or disagree with each other.


                    \subsubsection{ Evaluating Multipole Moments of Regular Black Holes using Thorne formalism} \label{regular2}
        
                    In this section, we will compute the multipole moments of Regular black holes Thorne formalism. 
                    \\
                    Expanding the metric, we get
        
                    \begin{eqnarray}
                        &g_{00} = - 1 + \frac{2M}{\sqrt{r^2 + l^2}} = - 1 + \frac{2M}{r} - \frac{M l^2}{r^3} + O\left(\frac{1}{r^4}\right) \nonumber \\
                        &g_{11} = \left(1 - \frac{2M}{\sqrt{r^ + l^2}} \right)^{-1} = 1 + \frac{2M}{r} + \frac{4M^2}{r^2} + \mathcal{O}\left(\frac{1}{r^3}\right) \nonumber \\
                        &g{0i} = 0 \nonumber \\
                        &g_{ii} =  1
                    \end{eqnarray}
                    here $i$ represents the spatial co-ordinates.
                    The spherical coordinates are the ACMC coordinates for Regular black holes. We get the moments as;
                    \begin{eqnarray}
                        &M_0 = - M \nonumber \\
                        &M_1 = 0,     M_2 = 0   
                    \end{eqnarray}
                    The higher order multipole moments are also zero since there is no $\theta$ dependence in the metric. Therefore, the multipole moments of Regular black holes Thorne formalism matches with the multipole moments of Regular black holes using improved twist form of Geroch-Hansen multipoles.

                \subsection{\label{app4} Multipole Moments of Kerr-Newman Black Hole using Thorne formalism}
        
                    \noindent
                    In this section, we evaluate the multipole moments defined by the Thorne formalism for the Kerr Newman metric. First, we note that the Kerr Newman metric in Boyer-Lindquist coordinates is given as follows: 
                    \begin{eqnarray}
                        ds^2  = -\left( \frac{\Delta - a^2 \sin^2{\theta}}{\Sigma}\right)dt^2 - 2a\sin^2{\theta} \left( \frac{2Mr -Q^2}{\Sigma}\right) dt d\phi \nonumber \\
                        + \frac{\Sigma}{\Delta} dr^2 + \Sigma d\theta^2 + \left( r^2 + a^2 +  \frac{2 M r - Q^2}{\Sigma}a^2 \sin^2{\theta}\right)\sin^2{\theta} d\phi^2 \nonumber
                    \end{eqnarray}
                    
                    \noindent
                    here $\Delta = r^2 -2Mr + a^2 + Q^2$ and $\Sigma = r^2 + a^2 \cos^2{\theta}$.
                    \\
                    To apply the Thorne formalism, we first need to expand the metric and check if the metric is already in ACMC coordinates. If not, we need to transform it into them. The non-zero metric coefficients of the Kerr metric are,
                    \begin{eqnarray}
                        g_{00} &=& \frac{-r^{2}+2 m r-a^{2} \cos ^{2} \theta - Q^2}{r^{2}+a^{2} \cos ^{2} \theta} = - 1 + \frac{2 m}{r} - \frac{Q^2}{R^2} \nonumber \\ 
                        &-& \frac{2 m a^{2} \cos ^{2} \theta}{r^{3}}+ \mathcal{O}\left(\frac{1}{r^{5}}\right)\\
                        g_{0 \phi} &=& -\frac{2 m a \sin \theta}{r^{2}} +\frac{Q^2asin\theta}{r^3} \nonumber + \frac{2 m a^{3} \sin \theta \cos ^{2} \theta}{r^{4}}\\
                        &+& \mathcal{O}\left(\frac{1}{r^{6}}\right)\\
                        g_{r r} &=& \frac{r^{2}+a^{2} \cos ^{2} \theta}{r^{2}-2 m r+a^{2} + Q^2} =1 + \frac{2 m}{r} + \frac{4 m^{2}-a^{2} \sin^{2}\theta}{r^2} \nonumber \\
                        &-& \frac{Q^2}{r^{2}} +\frac{8 m^{3}-2 m a^{2}\left(2-\cos ^{2} \theta\right)}{r^{3}}+ \mathcal{O}\left(\frac{1}{r^{4}}\right)\\
                        g_{\theta \theta}&=&  \frac{r^{2}+a^{2} \cos ^{2} \theta}{r^{2}}=1+\frac{a^{2}-a^{2} \sin ^{2} \theta}{r^{2}}\\
                        g_{\Phi \Phi} &=&  1+\frac{a^{2}}{r^{2}}+\frac{2 m a^{2} \sin ^{2} \theta}{r^{3}}+\frac{Q^4 a^4 \sin^2\theta}{r^4}+ {\mathcal{O}\left(\frac{1}{r^{5}}\right)}
                    \end{eqnarray}
                    The coefficient of $1 / r^{2}$ in $g_{00}$ is constant; hence the coordinates are mass centered. However, $g_{r r}$ and $g_{\theta \theta}$ metric elements contain $-a^{2} \sin ^{2} \theta / r^{2}$ at $\mathcal{O}\left(1 / r^{2}\right)$. Thus, the coordinates are ACMC-0; we shall transform to an ACMC-2 coordinate to determine the monopole, dipole, and quadrupole moments. 
                    \\
                    This can be accomplished by the transformation
                    \begin{eqnarray} \label{thorne transform}
                      &r=r^{\prime}+a^{2} \cos ^{2} \theta^{\prime} / 2 r^{\prime}, \theta=\theta^{\prime}-a^{2} \cos \theta^{\prime} \sin \theta^{\prime} / 2 r^{\prime 2}, \nonumber \\
                    &\phi=\phi^{\prime}, \quad t=t^{\prime},  
                    \end{eqnarray}
                    Applying the transformation leads to   
                    \begin{eqnarray}\label{KN 1}
                        g_{0^{\prime} 0^{\prime}}  & = & -1+\frac{2 m}{r^{\prime}}+\frac{Q^2}{r^{\prime2}}-\frac{3 m a^{2} \cos ^{2} \theta^{\prime}}{r^{\prime 3}} + \mathcal{O}\left(\frac{1}{r^{\prime 5}}\right) \nonumber \\
                         & = & -1+\frac{2 m}{r^{\prime}} +\frac{Q^2}{r^{\prime2}}-\frac{2 m a^{2}}{r^{\prime 3}}\left[P^{2}\left(\cos \theta^{\prime}\right) \right.\nonumber \\
                         & + & \left. \frac{1}{2} \right]+\mathcal{O}\left(\frac{1}{r^{\prime 5}}\right)
                    \end{eqnarray}
                    Now, we expand apply transformation and expand the $g_{03}$ component and get,
                    \begin{eqnarray}\label{KN 2}
                        g_{0^{\prime} \phi^{\prime}}  &=& -\frac{2 m a \sin \theta^{\prime}}{r^{\prime 2}}+\frac{Q^2 a \sin\theta}{r^{\prime3}} \nonumber \\
                        && + \frac{5 m a^{3} \sin \theta^{\prime} \cos ^{2} \theta^{\prime}}{r^{\prime 4}}+O\left(\frac{1}{r^{\prime 6}}\right) \nonumber \\
                        & = &\frac{2 m a P_{, \theta^{\prime}}^{1}}{r^{\prime 2}}+\frac{Q^2 a}{r^{\prime3}} P_{, \theta^{\prime}}-\frac{m a^{3}}{r^{\prime 4}}\left(P_{, \theta^{\prime}}^{1} \right. \\
                        && \left.+\frac{2}{3} P_{, \theta^{\prime}}^{3}\right)+O\left(\frac{1}{r^{\prime 6}}\right)
                    \end{eqnarray}
                    The other non-zero metric terms transform as
                    \begin{eqnarray}
                        g_{r^{\prime} r^{\prime}} =&1& +\frac{2 m}{r^{\prime}}+\frac{4 m^{2}-a^{2}- Q^2}{r^{\prime 2}} \nonumber \\
                        &+& \frac{8 m^{3}-4 m a^{2}-m a^{2} \cos ^{2} \theta^{\prime}}{r^{\prime 3}}+\mathcal{O}\left(\frac{1}{r^{\prime 4}}\right) \nonumber \\
                        g_{\theta^{\prime} \theta^{\prime}} = &1& + \frac{a^{2}}{r^{\prime 2}}+O\left(\frac{1}{r^{\prime 4}}\right)\nonumber \\
                        g_{\phi^{\prime} \phi^{\prime}} = &1& + \frac{a^{2}}{r^{\prime 2}}+\frac{2 m a^{2} \sin ^{2} \theta^{\prime}}{r^{\prime 3}}+O\left(\frac{1}{r^{\prime 4}}\right) 
                    \end{eqnarray}

                    \noindent
                    As we can see in these coordinates, no term $r^{-(\textit{l}+1)}$ has angular dependence on spherical harmonics higher than the order $\textit{l}$ till $\mathcal{O}(1/r^3)$ . Therefore, we can safely conclude that our coordinate is at least ACMC-2, and we can read monopole, dipole, and quadrupole moments.
                    \\
                    Hence, we compare eq. \eqref{KN 1} and eq. \eqref{KN 2} to eq. \eqref{2.2.1.a} and eq. \eqref{2.2.1.b} and monopole, dipole, and quadrupole moments are as follows.
                    \begin{eqnarray}
                        &M_0 = - M, \: M_1 = 0, \: M_2 = M a^2  \nonumber \\
                        &S_0 = 0, \: S_1 = M a, \: S_2 = 0
                    \end{eqnarray}
                    \noindent
                    The multi-polar structure found above is the same just like in the case of Kerr black holes. Therefore, the Thorne formalism doesn't predict any difference in the multipole moments due to the addition of a energy momentum tensor in Kerr black holes.

\nocite{*}
\bibliographystyle{ACM-Reference-Format}
\bibliography{apssamp}

\end{document}